\documentclass[a4paper]{IEEEtran}
\IEEEoverridecommandlockouts

\usepackage{cite}
\usepackage{amsmath,amssymb,amsfonts}
\usepackage{algorithmic}
\usepackage{graphicx}
\usepackage{textcomp}
\usepackage{xcolor}
\usepackage{mcite}
\usepackage{enumitem}
\usepackage{physics}
\usepackage{subcaption}
\usepackage{flushend}
\usepackage{tikz}
\usepackage{accents}
\usetikzlibrary{shapes.geometric, arrows}

\def\BibTeX{
{\rm B\kern-.05em{
\sc i\kern-.025em b
}\kern-.08em
T\kern-.1667em\lower.7ex
\hbox{E}\kern-.125emX}}

\usepackage{tikz}

\IEEEoverridecommandlockouts

\allowdisplaybreaks

\IEEEaftertitletext{\vspace{-0.5\baselineskip}}

\begin{document}
\bstctlcite{IEEEexample:BSTcontrol}

\title{
Reformulating dq Impedance Matrices via Pauli Decomposition for Root-Cause Analysis of Instabilities in Grid-Connected Converters
\thanks{
Josue Andino Bustamante, Milan Prodanovic and Javier Roldán-Pérez are with IMDEA Energy Institute, Madrid, Spain.
Javier Roldán-Pérez is also with Escuela Técnica Superior de Ingenieros Industriales, Universidad Politécnica de Madrid, Madrid, Spain. 
Corresponding author: Josue Andino (josue.andino@imdea.org).
Josue Andino is also a Ph.D. student at University of Alcalá de Henares.}
\thanks{
This work was supported by REDESFUERTES project (PID2022-142416OB-I00), financed by MICIU/AEI/10.13039/501100011033 and by FEDER, EU.
It was also supported by DYNAMIC COMPENSATION project (CPP2022-010120), also financed by MICIU/AEI/ 10.13039/501100011033, and by the European Union, NextGenerationEU/PRTR.}
}

\author{\IEEEauthorblockN{Josue Andino, \textit{Member, IEEE}, Milan Prodanovi\'c, \textit{Senior Member, IEEE}, and \\ Javier Roldán-Pérez, \textit{Senior Member, IEEE} \vspace{-0.2cm}}
}

\maketitle

\begin{abstract}
The increasing penetration of converter-interfaced generators in power systems has led to the adoption of impedance-based criteria as an alternative framework for assessing and ensuring stable integration.
However, when the impedance criterion is used, identifying the root cause of instabilities is generally more challenging compared to other approaches, such as modal analysis.
Moreover, the eigenvalues and characteristic equation used in the impedance criterion are non-linear functions, making it difficult to establish a clear relationship between impedance components and closed-loop stability.
To address this issue, this paper proposes the application of the Pauli decomposition to analyse dq impedance matrices and minor-loop equations.
By using this decomposition technique, the dq representation can be reformulated into a quaternion-like form, which has explicit algebraic relationships with the determinant, trace, eigenvalues, and characteristic equation.
Moreover, this decomposition enables systematic assessment of the influence of each impedance term in the system stability, thus facilitating finding the root-cause of instabilities.
The primary objective of this work is to develop the mathematical foundation of the Pauli decomposition and demonstrate its implications for root-cause analysis.
The theoretical contributions are validated using a case study consisting of a converter-interfaced generator connected to a weak grid that has been previously analysed in the literature using existing techniques.
The proposed Pauli decomposition provides an algebraic tool that enhances interpretability of impedance-based stability analysis and establishes a basis for further investigation of complex converter interactions.

\end{abstract}
\section{Introduction}
In recent years the number of renewable energy sources integrated into the grid via electronic power interfaces has increased significantly. 
These converters have several advantages compared to traditional synchronous generators, such as fast control action and application versatility~\cite{SU2020475}.

However, power electronic converters are known to produce interactions with other power converters or nearby generators~\cite{Hatziargyriou2021,cheng2021ibr}.
Unless they are properly damped, these interactions can lead to instabilities and are usually studied using small-signal stability analysis tools such as modal analysis~\cite{kundur1994power}.
Although the latter is a very effective tool, it strongly relies on analytical models that may not be available in realistic applications.
The impedance method represents an alternative tool for analysing  interactions in power systems dominated by converter-interfaced generators.
This tool is suitable for applications where analytical models are not available and only the frequency response of devices at the point of connection to the grid is required~\cite{Zhang2020,MCPE_SmallSignal,Pedra2024}.

The impedance criterion consists of calculating the eigenvalues of the minor loop, which is the product between the grid impedance and the converter admittance matrices~\cite{Sun2011,Amin2019,Fan2020}.
These matrices can be represented by a two-by-two real transfer matrix (in $dq$ coordinates), a two-by-two complex transfer matrix (in $pn$ sequence domain), or by a pair of complex transfer functions (CTFs)~\cite{Wang2018}.
While all representations contain the same information, each one has certain advantages and disadvantages for performing stability analyses.

One of the main drawbacks of the impedance criterion is the fact that determining the root-cause of instabilities is not straightforward,  even if analytical models are available.
There are two main reasons for that:
The first is that stability is usually assessed using the eigenvalues of impedance matrices that are difficult to interpret, and the second is related to the way impedance matrices are represented, which is not intuitive~\cite{Rygg2016,Amin2019}.
In addition, the calculation of the minor loop eigenvalues is usually seen as a black-box operator, i.e. the relationship between the input and the output data is unclear.
In fact, calculating the eigenvalues of two-by-two matrices involves the use of square roots, which makes it harder to interpret the results.
In the literature, there are two main alternatives to eigenvalues for stability assessment, namely the use of the characteristic equation (the determinant of the minor loop) and passivity indices.
However, in both cases, the interpretation of the results is based on the way the impedance is represented.
When $pn$ impedance matrices are used, CTFs, complex vectors, and their conjugates need to be operated, and this requires the use of non-linear operators. 
Also, for the case of $dq$-impedance matrices, the results are difficult to interpret, since they do not have a clear physical meaning.

The analysis of impedance matrices for the assessment of converter-driven interactions has been addressed in the literature in the past~\cite{10262032}.
One approach is to represent control loops as circuit elements, which is helpful for understanding the effect of controllers on the impedance matrix~\cite{Li2021}.
Yet, finding the relationship between the impedance terms and the stability properties is not straightforward.
Zhu \textit{et al.}~\cite{Zhu2022,Zhu2023} propose a method to calculate the participation of each term of the impedance in specific interactions, thus providing valuable information for the assessment of the root causes of interactions.
Still, this method depends on the analytical model of each device, and this information is not always available.
Another approach is to replace the products of the minor loop by additions~\cite{Li2020}.
This, however, requires the inversion of matrices and, therefore, significantly impedes the interpretation of the results.
Moutevelis~\textit{et al.}~\cite{DionysiosMoutevelis2024} design the controllers of power converters accounting for the stability margins derived from the application of the impedance criterion.
Yet, the tuning procedure is time consuming since the link between the control parameters and the stability properties should be assessed numerically. 
In this literature review, it has been shown that finding the root causes of instabilities using the impedance criterion is challenging, mainly because admittance and impedance matrices are difficult to interpret.

Several researchers are currently proposing the use of advanced mathematical tools such as quaternions and geometric algebra (GA) to model and study power system applications~\cite{MentiTCAS,MiltonGA,CastillaGA}. 
For example, Fang~\textit{et al.}~\cite{Fang2024} propose using quaternions to define the power, impedance, and admittance of three-phase systems.
This approach allows generalising three-phase electric quantities, yet stability aspects were not considered in the study.
These studies have also been performed using Park's and Clarke's transformation, in the GA framework~\cite{FGIL_GeometricAlgebra}. 
In contrast, Velasco~\textit{et al.}~\cite{García2024} propose using GA to analyse the stability of three-phase power converters.
This methodology allows one to transform a dynamic representation of a multi-input multi-output (MIMO) system into a single-input single-output (SISO) system, thus reducing its mathematical complexity.
Unfortunately, this methodology has not been developed for the application of the impedance criterion.
Apart from these works, there are several articles on GA for power electronics applications, but they mainly focus on the analysis of power flows in multiphase systems~\cite{Hitzer2024}.
A practical way of representing quaternions is Pauli matrices, since there is no need to use advanced mathematical operators~\cite{Tudor2010}. 
These matrices have relevant mathematical properties that are helpful in establishing links between the system description, eigenvalues, and determinants.
To the best of the authors knowledge, the application of Pauli decomposition to impedance matrices has not been previously explored in the literature.
Furthermore, based on the main findings of the literature review, this representation has great potential to help understanding interactions.

The primary contribution of this paper lies in formalizing the Pauli-algebra decomposition of impedance matrices and clarifying the structural meaning of its components.
This novel way of representing impedance matrices transforms them into quaternion-like structures, allowing the analysis of the impedance as a linear SISO system instead of a MIMO system.
Moreover, it is shown that the impedance quaternion has a direct link with stability criteria, such as eigenvalues and the characteristic equation.
As a consequence, Pauli decomposition becomes a helpful mathematical tool for finding the root cause of interactions.
In addition, the Pauli decomposition provides an interpretation for the terms of the impedance matrices and, therefore, simplifies understanding the way power electronic converters interact.
The theoretical results of the proposed methodology are verified using electromagnetic transient (EMT) simulations, performed in MATLAB/Simulink and its SimpowerSystems toolbox. 
A well-documented system from the literature is adopted as a benchmark for validation.
Experimental results using real hardware are intentionally omitted as the main findings are analytical and, moreover, the test system is taken from literature so the correctness of results can be simply assessed. 
The scripts and data needed to replicate the results of this work can be found in a public repository~\cite{pauliCode}.
The objective of this work is to establish the mathematical structure of the Pauli decomposition for impedance matrices and to clarify its algebraic and structural properties, not to provide a complete stability assessment methodology. 
The extension of this formulation to more complex systems and its use for systematic controller design remain subjects for future research.

The rest of the article is organised as follows.
The proposed Pauli decomposition of impedance matrices and its properties are described in Section~\ref{sec:decomp}.
Section~\ref{sec:stability} discusses the stability criteria using Pauli impedance decomposition.
Section~\ref{sec:root_cause} presents how the Pauli decomposition can be used to analyse the root-cause of interactions.
The case study is then presented, modelled and analysed in Section~\ref{sec:case}.
Finally, the conclusions are drawn in Section~\ref{sec:conclusions}.

\section{Pauli Impedance Decomposition}\label{sec:decomp}

\begin{figure*}[!t]
\centering
\begin{subfigure}[b]{0.31\linewidth}
\centering
\includegraphics[width=\linewidth]{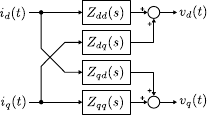}
\caption{}
\label{fig:dq_seq_imp}
\end{subfigure}
\hfill
\begin{subfigure}[b]{0.36\linewidth}  
\centering 
\includegraphics[width=1\linewidth]{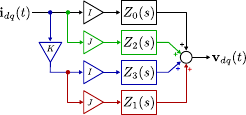}
\caption{}
\label{fig:pauli_imp}
\end{subfigure}
\hfill
\begin{subfigure}[b]{0.25\linewidth}  
\centering 
\includegraphics[width=\linewidth]{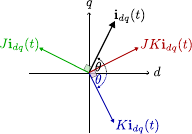}
\caption{}
\label{fig:current_vectors}
\end{subfigure}
\caption{(a) Conventional interpretation of the impedance matrix. (b) Pauli decomposed impedance (c) actions of the Pauli matrices on a vector. \vspace{-0.25cm}}
\label{fig:topologies}
\end{figure*}

The impedance of a system can be represented as a real transfer matrix in the $dq$ frame ($\mathbf{Z}_{dq}(s)$), a complex transfer matrix in the positive-negative sequence ($\mathbf{Z}_{pn}(s)$), or as a pair of complex transfer functions (CTFs) ($Z_{dq+}(s)$ and $Z_{dq-}(s)$)~\cite{Wang2018,Harnefors2020}.
For the case of CTFs, the system equations can be written as follows: 
\begin{equation}
\label{eqn:cmpximp}
v_{dq}(s) 
= 
Z_{dq+}(s)
i^{dq}(s) 
+
Z_{dq-}(s)
i_{dq}^*(s^*), 
\end{equation}
where the input current ($i_{dq}$) and the output voltage ($v_{dq}$) are complex variables, e.g., $i_{dq}(s)$~$=$~$i_d(s) $~$+$~$ji_q(s)$, and $j$ is the imaginary unit.  
The complex conjugate is marked using an asterisk $^*$, e.g., the complex conjugate of $i_{dq}$ is  $i_{dq}^*(s^*)$~$=$~$i_d(s)$~$-$~$ji_q(s)$.

As $Z_{dq+}(s)$ and $Z_{dq-}(s)$ are CTFs, they can also be expressed using their real and imaginary components:
\begin{align}
Z_{dq+}(s) 
&= 
Z_0(s) + jZ_2(s), \\ 
Z_{dq-}(s) 
&= 
Z_3(s) + jZ_1(s), 
\end{align}
where $Z_0(s)$, $Z_1(s)$, $Z_2(s)$, and $Z_3(s)$ are real-valued transfer functions (RTFs).
From now on, the variable $s$ will be omitted whenever possible.
Expression (\ref{eqn:cmpximp}) can be rewritten as:
\begin{equation}
v_d+jv_q 
= 
(Z_0+jZ_2)(i_d+ji_q)
+
(Z_3+jZ_1)(i_d-ji_q).
\label{eqn:expanded}
\end{equation}
By expanding (\ref{eqn:expanded}) and rewriting the result in vector form, the following result is obtained:
\begin{equation}
\label{eqn:explicit_form}
\underbrace{
\begin{bmatrix}
v_d \\ 
v_q 
\end{bmatrix}
}_{\displaystyle \mathbf{v}_{dq} }
= 
Z_0
\underbrace{
\begin{bmatrix}
i_d\\
i_q
\end{bmatrix}
}_{\displaystyle \mathbf{i}_{dq}} 
+ 
Z_2
\underbrace{
\begin{bmatrix}
-i_q\\
\phantom{-}i_d
\end{bmatrix}
}_{\displaystyle J\mathbf{i}_{dq} }
+ 
Z_3
\underbrace{
\begin{bmatrix}
\phantom{-}i_d\\-i_q
\end{bmatrix}
}_{\displaystyle K\mathbf{i}_{dq} } + 
Z_1
\underbrace{
\begin{bmatrix}
i_q\\i_d
\end{bmatrix}
}_{\displaystyle JK\mathbf{i}_{dq} },
\end{equation}
where $\mathbf{v}_{dq}$ and $\mathbf{i}_{dq}$ are real vectors representing the complex voltage ($v_{dq}$) and current ($i_{dq}$),  respectively.
$I$ is the $2\times 2$ identity matrix, and $J$, $K$, and $JK$ correspond to the real form of the standard Pauli matrices:

\begin{equation}
\label{eqn:pauli_matrices}
\mqty{
I\,\,\,
= 
\begin{bmatrix}
1 & \phantom{-}0 \\
0 & \phantom{-}1 
\end{bmatrix}, 
& \phantom{K}J
= 
\begin{bmatrix} 
0 & -1\\
1 & \phantom{-}0 
\end{bmatrix}, 
\\
K
= 
\begin{bmatrix} 
1 & \phantom{-}0 \\
0 & -1 
\end{bmatrix}, 
& JK
= 
\begin{bmatrix}
0 & \phantom{-}1
\\1 & \phantom{-}0 
\end{bmatrix}.
}
\end{equation}

All elements in (\ref{eqn:explicit_form}) can be merged together to calculate the impedance, producing the $dq$ impedance matrix:
\begin{align}
\label{eqn:pauli_impedance}
\mathbf{Z}_{dq}(s)
&=
Z_0(s)I + Z_2(s)J + Z_3(s)K + Z_1(s)JK.
\end{align}
Note that any $2\times 2$ complex matrix can be uniquely expressed as a linear combination of the Pauli matrices in which the coefficients are complex.
Since $Z_0(s)$, $Z_1(s)$, $Z_2(s)$, and $Z_3(s)$ are RTFs, the Pauli decomposition provides a complete representation of any $2\times 2$ impedance matrix.
The relationship between the conventional representation of the impedance matrix (Fig.~\ref{fig:dq_seq_imp}) and the proposed representation  (Fig.~\ref{fig:pauli_imp}) is:
\begin{align}
\label{eqn:matrix_rep}
\mathbf{Z}_{dq}(s)
=
\mqty[ Z_{dd} & Z_{dq} \\ Z_{qd} & Z_{qq} ]
=
\mqty[ Z_0+Z_3 & Z_1-Z_2 \\ Z_1+Z_2 & Z_0-Z_3 ].
\end{align}
Fig.~\ref{fig:dq_seq_imp} shows a graphical representation that can be used to understand the role of the elements of $\mathbf{Z}_{dq}(s)$.
It can be seen that each of the outputs ($v_d$ and $v_q$) is affected by the two inputs ($i_d$ and $i_d$), according to the value of the impedance terms. 
In contrast, by using the Pauli decomposition, the input vector is individually multiplied by the impedance terms $Z_0(s)$, $Z_1(s)$, $Z_2(s)$, and $Z_3(s)$, and then added. 
Also, it can be noticed that a diagonal and symmetrical impedance matrix will only consist of the term $Z_0$.
Meanwhile, the terms $Z_1$, $Z_2$, and $Z_3$ account for asymmetries and coupling terms.

From now on, $Z_0(s)$ is referred to as symmetrical impedance, as the sequence of the current will be the same as that of the voltage.
For example, if a positive-sequence current is injected in (\ref{eqn:explicit_form}), the effect of $Z_0(s)$ will only be a positive sequence voltage.
The term $Z_2(s)$ generates an output of the same sequence of the input, but with an additional phase shift of 90 degrees, due to the effect of $J$ (green arrow in Fig.~\ref{fig:current_vectors}).
For that reason, $Z_2(s)$ is named quadrature self-coupling impedance.
On the contrary, $Z_3(s)$ outputs a voltage in the opposite sequence than the input current.
For example, if a current with positive frequency is injected, $Z_3(s)$ will respond with a voltage at the mirror frequency (i.e., negative frequency).
This change of sequence is due to $K$ (blue arrow in Fig.~\ref{fig:current_vectors}), which is equivalent to the conjugation of a complex number.
Thus, $Z_3(s)$ will be referred as symmetrical cross-coupling impedance.
More details on mirror frequencies can be found in~\cite{Rygg2016}.
Finally, the effect of $Z_1(s)$ can be understood as the combined the effects of $Z_2(s)$ and $Z_3(s)$.
It outputs a voltage in the opposite sequence than the input current with an additional phase shift of 90$^\circ$ due to $JK$ (red arrow in Fig.~\ref{fig:current_vectors}).
Thus, $Z_1(s)$ is named quadrature cross-coupling impedance.

It is important to mention that the previous interpretation of the impedances $Z_0(s)$, $Z_1(s)$, $Z_2(s)$, and $Z_3(s)$ is valid not only in $dq$ coordinates, but also in positive-negative sequence ($\mathbf{Z}_{pn}(s)$) and even in $\alpha\beta$ coordinates ($\mathbf{Z}_{\alpha\beta}(s)$).
The relationship between the Pauli decomposition in $dq$ coordinates an other representations is described in more detail in Appendix~\ref{sec:rel}.
From now on, Pauli decomposition is always performed in $dq$ as all the impedance terms are RTFs, which are simpler to interpret.

\subsection{Modelling Example}\label{sec:freq}
In this section, the Pauli decomposition is used to model a simple series \textit{RL} circuit, in $dq$ coordinates.
First, the $\alpha\beta$ representation of the impedance is:
\begin{align}
Z_{\alpha\beta}^{RL}(s) 
= 
R + sL.
\end{align}
In $dq$ coordinates, the impedance matrix can be written as:
\begin{align}
\label{eqn:dq_mat_RL}
    \mathbf{Z}_{dq}^{RL}(s)
    =
    \mqty[ R+sL & -\omega_1L \\ \omega_1L & R+sL ], 
\end{align}
where $\omega_1$ is the synchronization frequency.
By applying Pauli decomposition to (\ref{eqn:dq_mat_RL}):
\begin{align}
\label{eqn:dq_pauli_RL}
    \mathbf{Z}_{dq}^{RL}(s)
    =
    (R+sL)I + (\omega_1L)J.
\end{align}
Indeed, the Pauli representation of the \textit{RL} circuit can be directly obtained by evaluating $s\rightarrow sI+J\omega_1$, as:
\begin{align}
\mathbf{Z}_{dq}^{RL}(s) 
= 
Z_{\alpha\beta}^{RL}
(sI+J\omega_1).
\end{align}
The evaluation at $s\rightarrow sI+J\omega_1$ will be known, from now on, as the ``matrix equivalent of frequency shifting''.

\vspace{-0.1cm}
\section{Pauli Impedance Quaternion}\label{sec:quaternion}
\vspace{-0.1cm}
\subsection{Definition of the Minor Loop}
Except in particular cases, the stability of an interconnected system consisting of an impedance ($\mathbf{Z}_{dq}(s)$) and an admittance ($\mathbf{Y}_{dq}(s)$) matrix can be analysed by means of the minor loop:
\begin{align}
\label{eqn:prod_mat}
\mathbf{L}_{dq}(s) 
&=
\mathbf{Z}_{dq}(s) \mathbf{Y}_{dq}(s),
\end{align}
with
\begin{align}
\label{eqn:pauli_admittance}
\mathbf{Y}_{dq}(s)
&=
Y_0(s)I+Y_2(s)J + Y_3(s)K+Y_1(s)JK.
\end{align}
Since $\mathbf{L}_{dq}(s)$ is a $2\times 2$ complex matrix, it can also be represented using Pauli decomposition:
\begin{align}
\label{eqn:L_terms}
\mathbf{L}_{dq}(s) 
&=
L_0(s)I+L_2(s)J 
+ 
L_3(s)K+L_1(s)JK.
\end{align}
The relationship between the components of the minor loop ($L_0$--$L_3$) and the impedance and admittance components ($Z_0$--$Z_3$ and $Y_0$--$Y_3$) can be deducted by multiplying the matrices.
For that purpure, it is of interest to understand the meaning of matrices $I$, $J$, $K$, and $JK$, when multiplied between them.

\vspace{-0.4cm}
\subsection{Properties of Pauli Matrices}
When $\mathbf{Z}_{dq}(s)$ and $\mathbf{Y}_{dq}(s)$ are multiplied, there are products between different Pauli matrices.
Some relevant properties of the Pauli matrices needed to understand the product between these two elements: 
\begin{enumerate}
\item The product between $J$ and $K$ anti-commutes, so $JK$~$=$~$ -KJ$.
The way of interpreting multiplication of Pauli matrices is the application of two transformations to the input vector.
\item The square of $K$ and $JK$ equals the identity ($I$). 
The square of $J$ is also $I$, with a minus sign ($-I$).
The following relation can be written:
\begin{align}
\label{eqn:square}
K^2 
= 
(JK)^2 
=
-J^2 
=
I.
\end{align}
The interpretation squared Pauli matrices is the application of a transformation on the input vector, two times.
For example, if a vector $\mathbf{i}_{dq}$ is reflected twice, the output will be the original vector (i.e., $K^2\mathbf{i}_{dq}$~$=$~$\mathbf{i}_{dq}$).
The same happens when swapping the elements of a vector two times: $(JK)^2\mathbf{i}_{dq}$~$=$~$\mathbf{i}_{dq}$.
Also, if a vector $\mathbf{i}_{dq}$ is rotated $90^\circ$ two times, the output will be the same vector with a phase shift of $180^\circ$: $J^2\mathbf{i}_{dq}$~$=$~$-\mathbf{i}_{dq}$.
\end{enumerate}

\vspace{-0.4cm}
\subsection{Pauli Quaternion Representation}
The Pauli impedance matrix presented in (\ref{eqn:pauli_impedance}) can be represented as a quaternion or a complexified quaternion (commonly knows as biquaternion).
For simplicity, the quaternion representation has been chosen in this work.
This representation helps in understanding some of the developments, and it will be briefly described.
Further information about quaternions and biquaternions can be found in \cite{Sangwine2011,Tudor2010}. 

From now on, $z(s)$ denotes a Pauli-quaternion representation of a $dq$ impedance matrix $\mathbf{Z}_{dq}(s)$, while its uppercase version, $Z_n(s)$, represents its impedance component: 
\begin{align}
\label{eqn:quaternion}
z(s)
&=
Z_0 + Z_1\sigma_1 - jZ_2\sigma_2 + Z_3\sigma_3
= 
Z_0
+ 
\vec{Z}\cdot\vec{\sigma}, 
\end{align}
where $JK$~$=$~$\sigma_1$, $J$~$=$~$-j\sigma_2$, and $K$~$=$~$\sigma_3$.
$Z_0$ is the scalar component, and $\vec{Z}$~$=$~$[Z_1, -jZ_2, Z_3]$ the vector component, that will be called coupling vector.
The vector $\vec{\sigma}=[\sigma_1, \sigma_2, \sigma_3 ]$ will be called ``Pauli vector'', and the dot in $\vec{Z}\cdot\vec{\sigma}$ refers to the dot product. 
The elements of $\vec{\sigma}$ can have a geometric interpretation, but this is not discussed here for simplicity (see~\cite{Zhang2017} for details).

One advantage of writing the impedance as a quaternion is that mathematical operations (such as multiplications) are already defined in the Pauli Algebra~\cite{Tudor2010}. 
In particular, the product between two Pauli quaternions can be written as:
\begin{align}
\label{eqn:prod_rule}
L(s) 
= 
z(s)y(s) 
= 
L_0 
+ 
\vec{L}\cdot\vec{\sigma},
\end{align}
with
\begin{align}
L_0 
&= 
Z_0Y_0 +  \vec{Z}\cdot\vec{Y}, \label{eqn:scalar_term}
\\
\vec{L} 
&= 
Z_0\vec{Y} 
+ 
Y_0\vec{Z} 
+
j(\vec{Z}\times\vec{Y}), \label{eqn:vector_term}
\end{align}
where $L(s)$ and $y(s)$ are the quaternion representation of the minor loop and the admittance, respectively.
Expressions $\vec{Z}\cdot\vec{Y}$ and $\vec{Z}\times\vec{Y}$ represent the dot and  cross product of the vectors involved.

Although representations based on Pauli matrices (in (\ref{eqn:pauli_impedance})) and quaternions (in (\ref{eqn:quaternion})) are equivalent, in this work, the former will be used for modelling purposes as results are easier to interpret.
Meanwhile, quaternion representation will be used to analyse stability, as mathematical manipulation is simpler.
\vspace{-0.4cm}

\subsection{Relevant Mathematical Functions}
\label{sec:semi-norm}
In this section, some mathematical tools needed to understand the rest of the paper are discussed. 
In particular, the semi-norm, the determinant and the dot product are explained.
\subsubsection{Semi-Norm and Determinant}
The norm of a quaternion is typically a real non-negative number, however, for complex quaternions, the norm is a complex number.
It is, therefore, commonly called ``semi-norm''.
The semi-norm of a quaternion is noted as $\norm{\cdot}$, and it is calculated as follows:
\begin{align}
\norm{z(s)}^2
:=&\,
Z_0^2-\vec{Z}\cdot\vec{Z}
=
Z_0^2 - Z_1^2 + Z_2^2 - Z_3^2.
\label{eqn:semi_norm}
\end{align}
The semi-norm is always an RTF (in $dq$) since $Z_0(s)$, $Z_1(s)$, $Z_2(s)$, and $Z_3(s)$ are also RTFs.
A property of the semi-norm that will be used later in the paper is:
\begin{align}
\label{eqn:det_comp}
\norm{z(s)}^2 
= 
\det{\vb*{Z}_{dq}(s)} = Z_{dd}Z_{qq} - Z_{dq}Z_{qd}.
\end{align}
A relevant aspect of Pauli decomposition is that the determinant is not longer a product of the matrix elements (as in (\ref{eqn:det_comp})), but a sum of the squared components (as in (\ref{eqn:semi_norm})).
This will be helpful in findings the root cause of instabilities.
Finally, a property that will be used for calculating the minor loop is:
\begin{align}
\norm{L(s)}^2
&=
\norm{z(s)y(s)}^2
=
\norm{z(s)}^2 \norm{y(s)}^2
\end{align}
\subsubsection{Dot Product}
The dot product of two complex quaternions ($z(s)$ and $y(s)$) is a complex number (except in very specific cases~\cite{Sangwine2011}):
\begin{align}
\label{eqn:dot_prod}
\langle z, y \rangle
:=&\,
Z_0Y_0 + \vec{Z}\cdot\vec{Y} \nonumber
\\=&\,
Z_0Y_0 + (Z_1)(Y_1) + (-jZ_2)(-jY_2) + (Z_3)(Y_3) \nonumber
\\=&\,
Z_0Y_0 + Z_1Y_1 - Z_2Y_2 + Z_3Y_3.
\end{align}
\vspace{-0.5cm}

\section{Stability Criteria}
\label{sec:stability}
In the conventional impedance criterion, closed-loop stability is usually assessed by analysing the eigenvalues of the minor loop or else the characteristic equation $(\chi(s))$.
In this work, the characteristic equation is used:
\begin{align}
\label{eqn:Ls}
\chi(s) 
= 
\det{ I 
+ 
\mathbf{L}_{dq}(s)}.
\end{align}
This expression can also be written by using the semi-norm:
\begin{align}
\chi(s) 
=
\norm{ 1 + L(s) }^2,
\end{align}
which can be expanded, as in (\ref{eqn:semi_norm}), yielding:
\begin{align}
\label{eqn:Ls_expanded}
\chi(s)
&=
(1+L_0)^2 - L_1^2 + L_2^2 - L_3^2 \nonumber
\\&=
1+2L_0 + L_0^2 - L_1^2 + L_2^2 - L_3^2 \nonumber
\\&=
1 + 2L_0(s) + \norm{L(s)}^2.
\end{align}
The mathematical properties of the semi-norm and the dot product presented in Section~\ref{sec:semi-norm} can be applied now to further simplify (\ref{eqn:Ls_expanded}), leading to:  
\begin{align}
\label{eqn:ls_pauli}
\chi(s) 
= 
1 
+ 
2\langle z, y \rangle 
+ 
\norm{z}^2\norm{y}^2.
\end{align}
Expanding (\ref{eqn:ls_pauli}), the details of the characteristic equation can be found:
\begin{align}
\label{eqn:Ls_full}
\chi(s) = 1 
&+
2(Z_0Y_0) + 2(Z_1Y_1) - 2(Z_2Y_2) + 2(Z_3Y_3) \nonumber
\\&
+ (Z_0Y_0)^2 - (Z_0Y_1)^2 + (Z_0Y_2)^2 - (Z_0Y_3)^2 \nonumber
\\&
- (Z_1Y_0)^2 + (Z_1Y_1)^2 - (Z_1Y_2)^2 + (Z_1Y_3)^2 \nonumber
\\&
+ (Z_2Y_0)^2 - (Z_2Y_1)^2 + (Z_2Y_2)^2 - (Z_2Y_3)^2 \nonumber
\\&
- (Z_3Y_0)^2 + (Z_3Y_1)^2 - (Z_3Y_2)^2 + (Z_3Y_3)^2.
\end{align}
It can be seen that $\chi(s)$ depends on the addition of impedance-admittance pairs, i.e., $Z_n(s)Y_m(s)$ with $n,m\in\{0,1,2,3\}$.
This expression will be later analysed to try to explain the root-case of interactions of interconnected elements. 

\vspace{-0.1cm}
\section{Root-Cause Analysis}
\label{sec:root_cause}
\vspace{-0.1cm}
\subsection{Contribution of Pairs to Stability}
\begin{figure}[!t]
\centering
\includegraphics[width=0.8\linewidth]{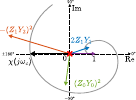}
\caption{Example of a Nyquist plot of $\chi(j\omega)$ along with three impedance-admittance pairs.}
\vspace{-0.2cm}
\label{fig:nyquist_example}
\end{figure}

Fig.~\ref{fig:nyquist_example} shows an example of a Nyquist plot of $\chi(j\omega)$ along with three relevant impedance-admittance pairs.
In this example, it is assumed the elements to be interconnected (with impedance and admittance matrices ($\mathbf{Z}_{dq}(s)$ and $\mathbf{Y}_{dq}(s)$) are stable when connected to ideal current and voltage sources.
Thus, the closed-loop system is unstable because $\chi(j\omega)$ encircles the origin.
The selected pairs are evaluated at the critical frequency ($j\omega_c$), which is the frequency at which the phase of $\chi(j\omega)$ equals $\pm 180^\circ$ (black arrow).
Observe that the pair $Z_1Y_1$ (blue arrow) is improving the system stability since it is pointing to the positive side of the real axis in the complex plane (at $j\omega_c$).
In contrast, the pair $Z_1Y_2$ (orange arrow) is pointing to left and, as a consequence, deteriorates the stability margins.
Finally, the pair $Z_0Y_0$ (green arrow) is almost perpendicular to the real axis.
Therefore, it can impact stability on a positive or negative way, depending on the case.
In this example, $Z_0Y_0$ and $Z_1Y_2$  cooperate to make $\chi(j\omega_c)$ encircle the origin.
Thus, it can be said that $Z_0Y_0$ and $Z_1Y_2$ are deteriorating the system stability.

\vspace{-0.4cm}
\subsection{Combined Contribution to Stability}
\label{sec:total_contributions}
To find the root cause of instabilities, the effect of all the pairs in (\ref{eqn:Ls_full}) must be analysed.
This process is tedious, as there are twenty cases that must be analysed (see the terms in (\ref{eqn:Ls_full})).
Moreover, the pairs $Z_nY_m$ for $n=m$ appear two times in (\ref{eqn:Ls_full})), further increasing the effect of any term on the system stability.

Commonly, only the admittance ($y(s)$) of the device to be connected to the grid ($z(s)$) can be modified.
Therefore, in order to design (or redesign) controllers, it is of interest to find what components of $y(s)$ are affecting at the conflicting frequencies.
Collecting all the components of $y(s)$, $\chi(s)$ can be written as follows:
\begin{align}
\label{ec.xi.func.l}
\chi(j\omega) 
= 
1 + \ell_0(j\omega) 
+ 
\ell_1(j\omega) 
+ 
\ell_2(j\omega) 
+ 
\ell_3(j\omega),
\end{align}
with
\begin{align}
\ell_0(j\omega) = \phantom{-}2Z_0(j\omega)Y_0(j\omega) + Y_0^2(j\omega)\norm{z(j\omega)}^2 \label{eqn:ell0},\\
\ell_1(j\omega) = \phantom{-}2Z_1(j\omega)Y_1(j\omega) - Y_1^2(j\omega)\norm{z(j\omega)}^2, \label{eqn:ell1}\\
\ell_2(j\omega) = -2Z_2(j\omega)Y_2(j\omega) + Y_2^2(j\omega)\norm{z(j\omega)}^2,\label{eqn:ell2}\\
\ell_3(j\omega) = \phantom{-}2Z_3(j\omega)Y_3(j\omega) - Y_3^2(j\omega)\norm{z(j\omega)}^2. \label{eqn:ell3}
\end{align}
In (\ref{ec.xi.func.l}), it can be observed that $\ell_m(j\omega)$ captures the contribution of the admittance component $Y_m(j\omega)$ on $\chi(j\omega)$.
To calculate (\ref{eqn:ell0})-(\ref{eqn:ell3}), only frequency responses are needed.
However, to fully understand the corrective actions needed to stabilise a system, a deeper look into the elements of $y(s)$ is needed.
Among the possible mitigation strategies, retuning of existing controllers can be effective when their parameters are accessible.
Alternatively, additional control loops can be included to compensate the phase or magnitude of the conflicting components on $y(s)$.
If models are available, the conflicting component $Y_m(s)$ can be traced back to its origin and subsequently addressed.
In the absence of such models, the decomposition itself can still provide qualitative insight into the likely source of instability.
For instance, the predominance of even components (e.g., $Y_0$ and $Y_2$), may indicate that the instability is primarily associated with passive elements, such as inductances or capacitances.
Conversely, predominance of odd components (e.g., $Y_3$ or $Y_1$) may indicate that the source is related with synchronisation loops, phase-locked loops (PLLs), voltage controllers, among others.
A systematic investigation of the relationship between impedance/admittance components and their physical interpretation in black-box scenarios remains an open research direction and will be addressed in future work.
The systematic design of compensators to stabilise grid-connected power electronic converters is beyond the scope of this work.

\section{Case Study}\label{sec:case}

\begin{figure*}[!t]
\centering
\includegraphics[width=0.8\linewidth]{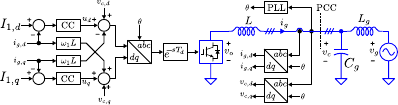}
\vspace{-0.1cm}
\caption{Electrical and control diagram of a power converter connected to the grid.}
\vspace{-0.1cm}
\label{fig:main_scheme}
\end{figure*}

\begin{table}[!b]
\centering
\begin{tabular}{c|ll}
\hline
Grid & Nominal Voltage ($V_1$) & 326~V \\
& Nominal frequency ($\omega_1$) & 50~Hz \\
& Parallel capacitance ($C_g$) & 20~$\mu$F \\
& Series inductance ($L_g$) & 5~mH \\
\hline
AC/DC & Output filter ($L$) & 3~mH \\
Converter & Delay time ($T_d$) & 150~$\mu$s \\
& Nominal $d$-axis current ($I_{1,d}$) & 15~A \\
& Nominal $q$-axis current ($I_{1,q}$) & 0~A \\
& Nominal $d$-axis voltage ($U_{1,d}$) & -0.88~V \\
& Nominal $q$-axis voltage ($U_{1,q}$) & 15.38~V \\
\hline
Current & Proportional gain ($K_{p-cc}$) & 16~$\Omega$ \\
Controller & Integral gain ($K_{i-cc}$) & 600~$\Omega$/s \\
\hline
Phase-locked & Proportional gain ($K_{p-pll}$) & 18.07~rad/s \\
loop Controller & Integral gain ($K_{i-pll}$) & 27708~rad/s$^2$ \\
\hline
\end{tabular}
\caption{Parameters of the case study.}
\label{tab:params}
\end{table}

\subsection{System Description}\label{sec:description}
Fig.~\ref{fig:main_scheme} shows the system studied in this work, which consists of a converter-interfaced generator connected to a weak grid.
This case study is similar to that analysed in~\cite{Wang2018}.
In the original case study, three scenarios are proposed and one of them is unstable due to the action of the PLL.
In this paper, the same unstable scenario is taken as the starting point.
However, it is assumed that the root cause of the instability is unknown and has to be found.
It is also assumed that the impedance of the grid ($\mathbf{Z}_{g}$) is fixed.
The grid consists of an ideal voltage source ($v_g$), with amplitude $V_1$ and frequency $\omega_1$, a series inductor ($L_g$) and a parallel capacitor ($C_g$).
The power converter consists of a three-phase inverter that generates the voltage $v_o$, and an output filter ($L$).
The control algorithm is based on a conventional current space-vector PI controller with decoupling terms and grid-voltage feed-forward.  
The computation delay is modelled by using a constant delay $T_d$ ($e^{-sT_d}$).
The controller includes a PLL to align the synchronous reference frame of the power converter with that of the grid voltage space vector.
The main objective of the power converter is the injection of constant currents ($I_{1,d}$ and $I_{1,q}$) into the grid.
Table~\ref{tab:params} shows the values of the parameters used in this work.

The selected case study is intentionally simple and well-documented in the literature, as the objective is to illustrate the properties of the Pauli decomposition rather than to provide an exhaustive system-level validation.
The proposed decomposition is algebraically general for any $2\times 2$ impedance representation and can be extended to more complex converter interactions, which will be addressed in future studies.

From now on, the subscript $dq$ will be also omitted since all the matrices are described in this frame.
Also, the modelling is done by using matrix representation, as in (\ref{eqn:pauli_impedance}).
The admittance and impedance models presented in this work have been validated via MATLAB/Simulink frequency response estimator tool.
Only relevant parts of the model are shown, while the full model is provided in the following repository:~\cite{pauliCode}.

\subsection{Test Case Admittance Modelling }
\label{sec:admittance_modelling}

\begin{figure}[!b]
\centering
\includegraphics[width=0.99\linewidth]{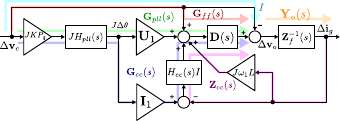}
\vspace{-0.35cm}
\caption{Admittance model of the power electronic converter.}
\vspace{-0.15cm}
\label{fig:small_signal_model}
\end{figure}

The converter admittance ($\mathbf{Y}_c$) is a transfer matrix that models the system from $\Delta \mathbf{v}_c$ to $\Delta \mathbf{i}_g$ (see Fig.~\ref{fig:small_signal_model}).
All the elements of the converter-interfaced generator have been modelled individually and then merged together.
The transfer matrices $\mathbf{G}_{ff}(s)$, $\mathbf{G}_{pll}(s)$, and $\mathbf{G}_{cc}(s)$ go from $\Delta\mathbf{v}_c$ to $\Delta\mathbf{v}_o$.
The impedance of the current controller ($\mathbf{Z}_{cc}(s)$) goes from $\Delta\mathbf{i}_g$ to $\Delta\mathbf{v}_o$. 
Finally, the equivalent output admittance ($\mathbf{Y}_o(s)$) goes from $\Delta\mathbf{v}_o$ to $\Delta\mathbf{i}_g$.
All these elements are connected in series with the delay, $\mathbf{D}(s)$, and the filter of the power electronic converter ($\mathbf{Z}_f(s)$).
The admittance of the device is calculated merging all the sub-blocks, yielding:
\begin{align}
\mathbf{Y}_c(s)
&= 
( \mathbf{Z}_f + \mathbf{Z}_{cc} )^{-1}( -I + \mathbf{G}_{ff} + \mathbf{G}_{pll} + \mathbf{G}_{cc} ) = \nonumber
\\&=
-\mathbf{Y}_{o} + \mathbf{Y}_{ff} + \mathbf{Y}_{pll} + \mathbf{Y}_{cc}, 
\label{eqn:yc_sum}
\end{align}
where $\mathbf{Y}_{ff}(s)$, $\mathbf{Y}_{pll}(s)$, and $\mathbf{Y}_{cc}(s)$ are the admittances produced by the feed-forward (FF) action, the PLL, and the CC,  respectively. 
The analytical models of the transfer functions are listed here for completeness:
\begin{align}
\mathbf{Z}_f(s) 
&= (R+sL)I + J\omega_1L,
\label{eqn:zf} \\
\mathbf{D}(s) 
&= 
e^{-(sI+J\omega_1)T_d} 
= 
e^{-J\omega_1T_d} e^{-sT_d},
\label{eqn:delay}\\
\mathbf{Z}_{cc}(s) 
&= 
\mathbf{D}(s) \qty[\, H_{cc}(s)I - J\omega_1L \,], 
\label{eqn:zcc} \\
\mathbf{G}_{pll}(s) 
&=
\mathbf{D}(s) \; \mathbf{U}_1 \; J \; H_{pll}(s) \; JK \; P_q,
\label{eqn:gpll} \\
\mathbf{G}_{cc}(s) 
&= 
\mathbf{D}(s) \; H_{cc}(s) \; I \; \mathbf{I}_1 \; J \; H_{pll}(s) \; JK \; P_q,
\label{eqn:gcc}\\
\mathbf{G}_{ff}(s) 
&= 
\mathbf{D}(s), 
\label{eqn:gff}
\end{align}
where $\mathbf{U}_1$ and $\mathbf{I}_1$ are the matrix representation of the steady-state voltage produced by CC and current injected to the grid respectively, i.e., $\mathbf{U}_1$~$=$~$U_{1,d}I$~$+$~$J U_{1,q}$, and $\mathbf{I}_1$~$=$~$I_{1,d}I$~$+$~$J I_{1,q}$.
$P_q$ is the projection operator on the $q$ axis, see Appendix~\ref{sec:reim}.
$\mathbf{Z}_f(s)$ and $\mathbf{D}(s)$ are obtained by applying a frequency shift to $R+sL$ and $e^{-sT_d}$, respectively ($s$~$\rightarrow$~$sI$~$+$~$J\omega_1$) (see Section~\ref{sec:freq}).
The transfer function of current controller model is:
\begin{align}
    H_{cc}(s)=K_{p-cc} + \dfrac{K_{i-cc}}{s}.
\end{align}
Finally, the transfer function of the PLL ($H_{pll}(s)$) is represented by its linearised model as in~\cite{Wang2018}:
\begin{equation}
H_{pll}(s) 
=
\dfrac{ \Delta \theta (s) }{ \Delta v_{c,q} (s) }
= 
\dfrac{ K_{p-pll}\;s + K_{i-pll} }{ s^2 + V_1 K_{p-pll}\;s + V_1K_{i-pll} }.
\end{equation}
\subsection{Grid Impedance Modelling}\label{sec:impedance_modelling}
The $dq$ impedance model of the grid ($\mathbf{Z}_{g}(s)$) can be obtained by applying the frequency shifting to the impedance of the grid in $\alpha\beta$ (see Section~\ref{sec:freq}):
\begin{align}
\mathbf{Z}_{g}(s) 
&= 
\eval*{ \qty[\, 
(R_g+sL_g)^{-1} + sC_g \,
]^{-1} }_{s\rightarrow 
sI+J\omega_1}.
\end{align}

\subsection{Closed-Loop Stability Assessment}

\begin{figure}[!t]
\centering
\includegraphics[width=0.99\linewidth]{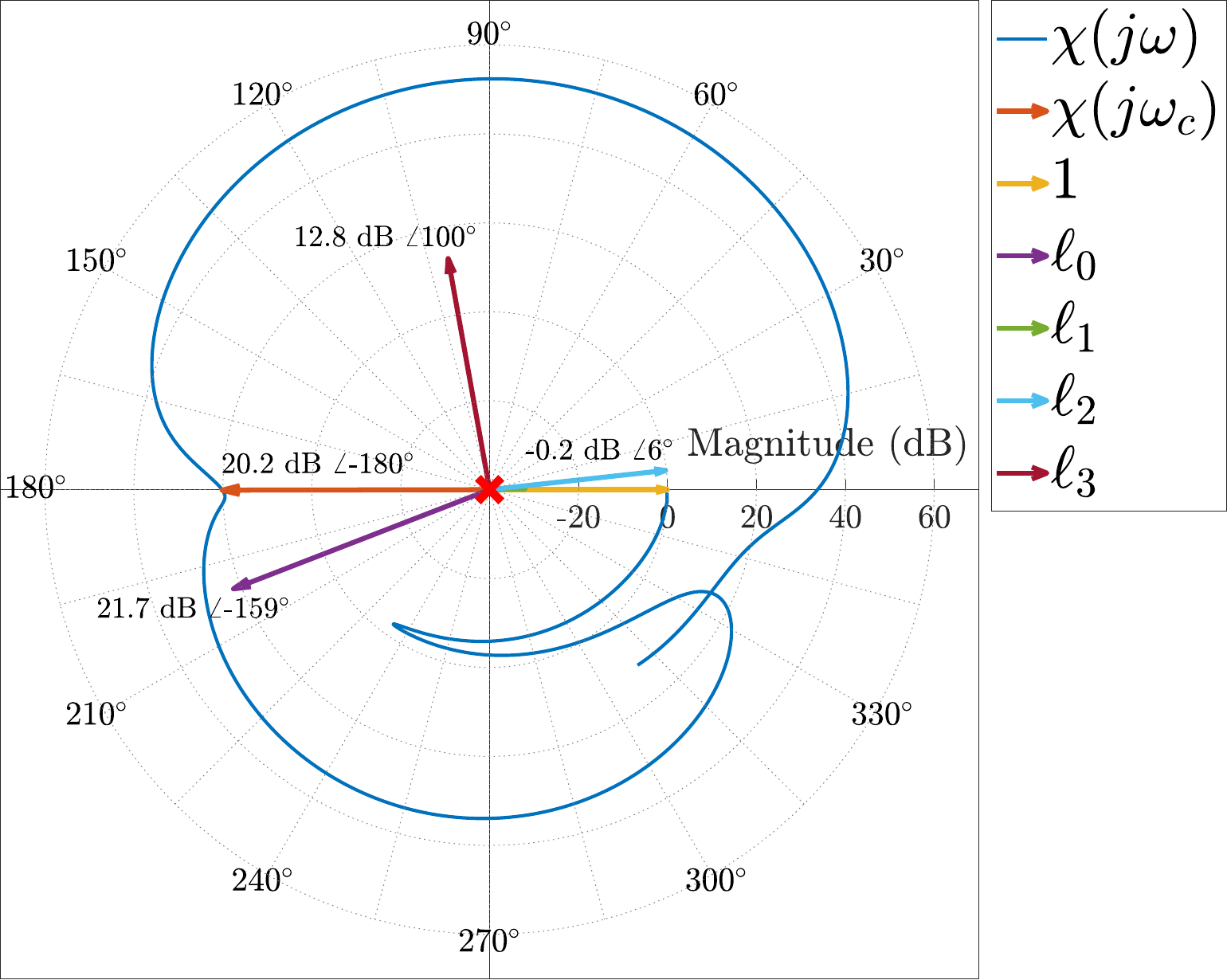}
\caption{Logarithmic Nyquist plot of $\chi(j\omega)$, for the unstable case.}
\label{fig:dignity_chart}
\end{figure}

\begin{figure}[!t]
\centering
\begin{subfigure}{0.75\linewidth}
\centering
\includegraphics[width=\linewidth]{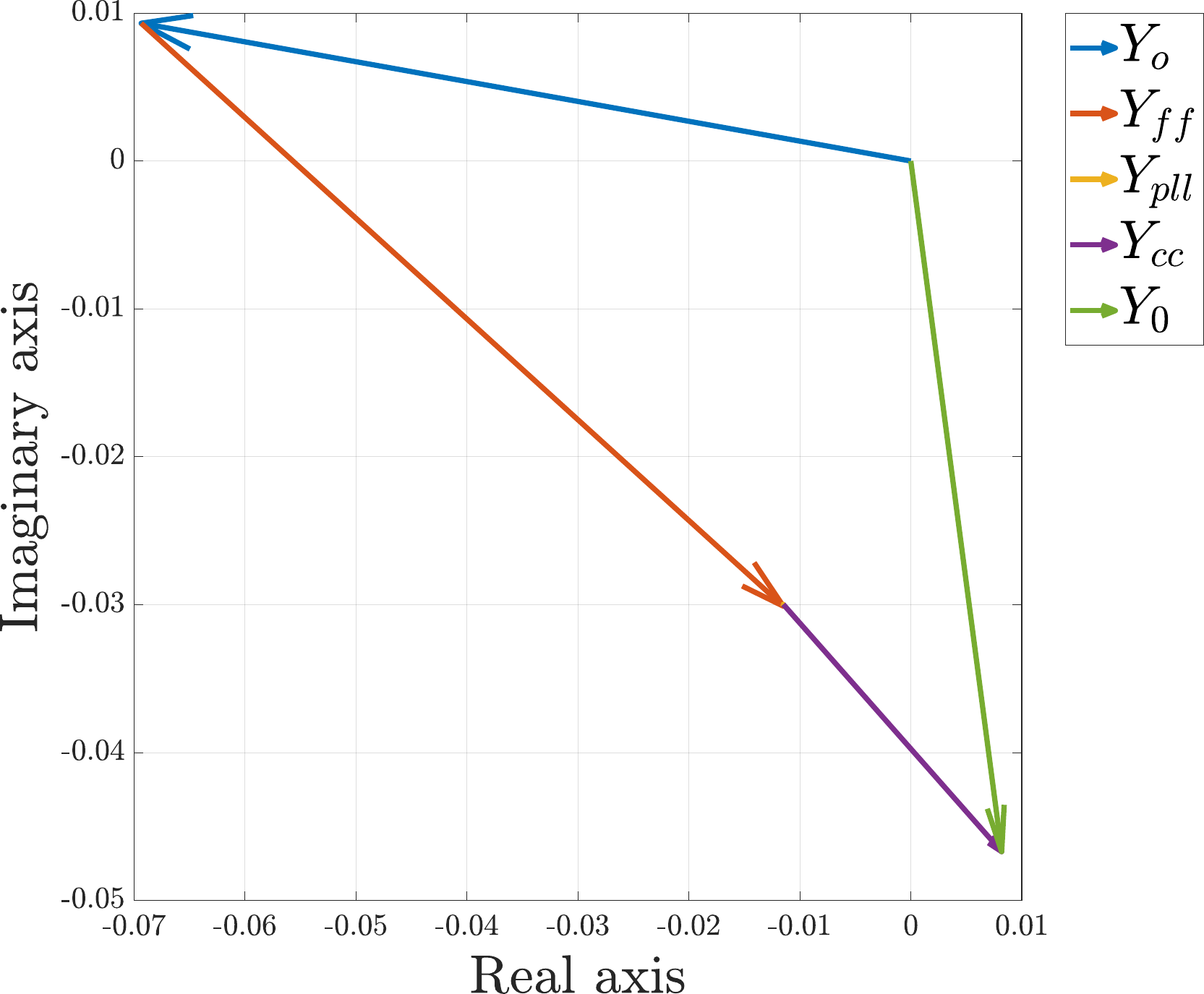}
\caption{Components of $Y_0(j\omega_c)$}
\label{fig:y0_components}
\end{subfigure}
\hfill
\begin{subfigure}{0.77\linewidth}
\centering
\includegraphics[width=\linewidth]{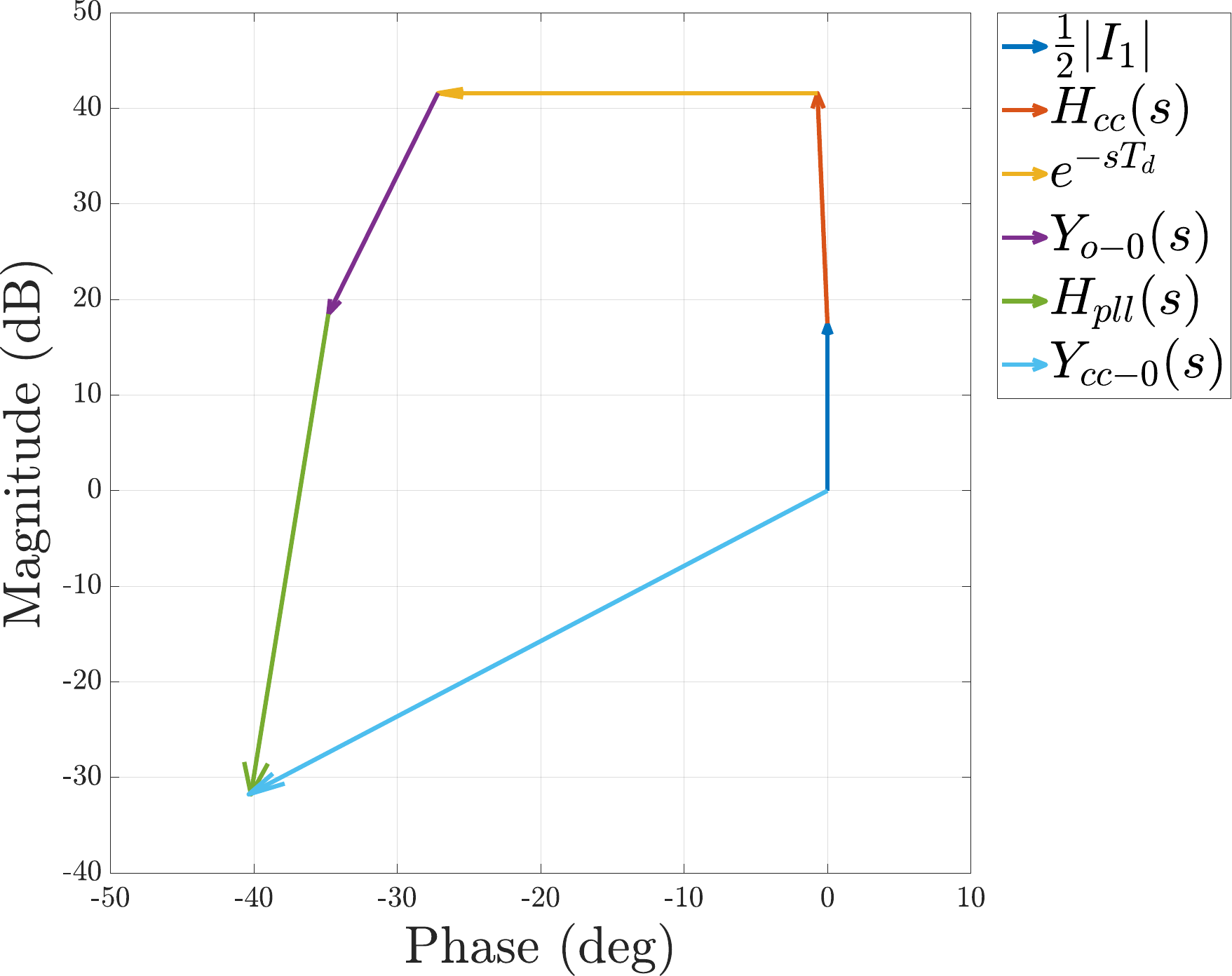}
\caption{Components of $Y_{cc-0}(j\omega_c)$}
\label{fig:cc0_components}
\end{subfigure}
\caption{Analysis of the components of the converter admittance.}
\label{fig:components}
\end{figure}

\begin{figure}[!b]
\centering
\includegraphics[width=\linewidth]{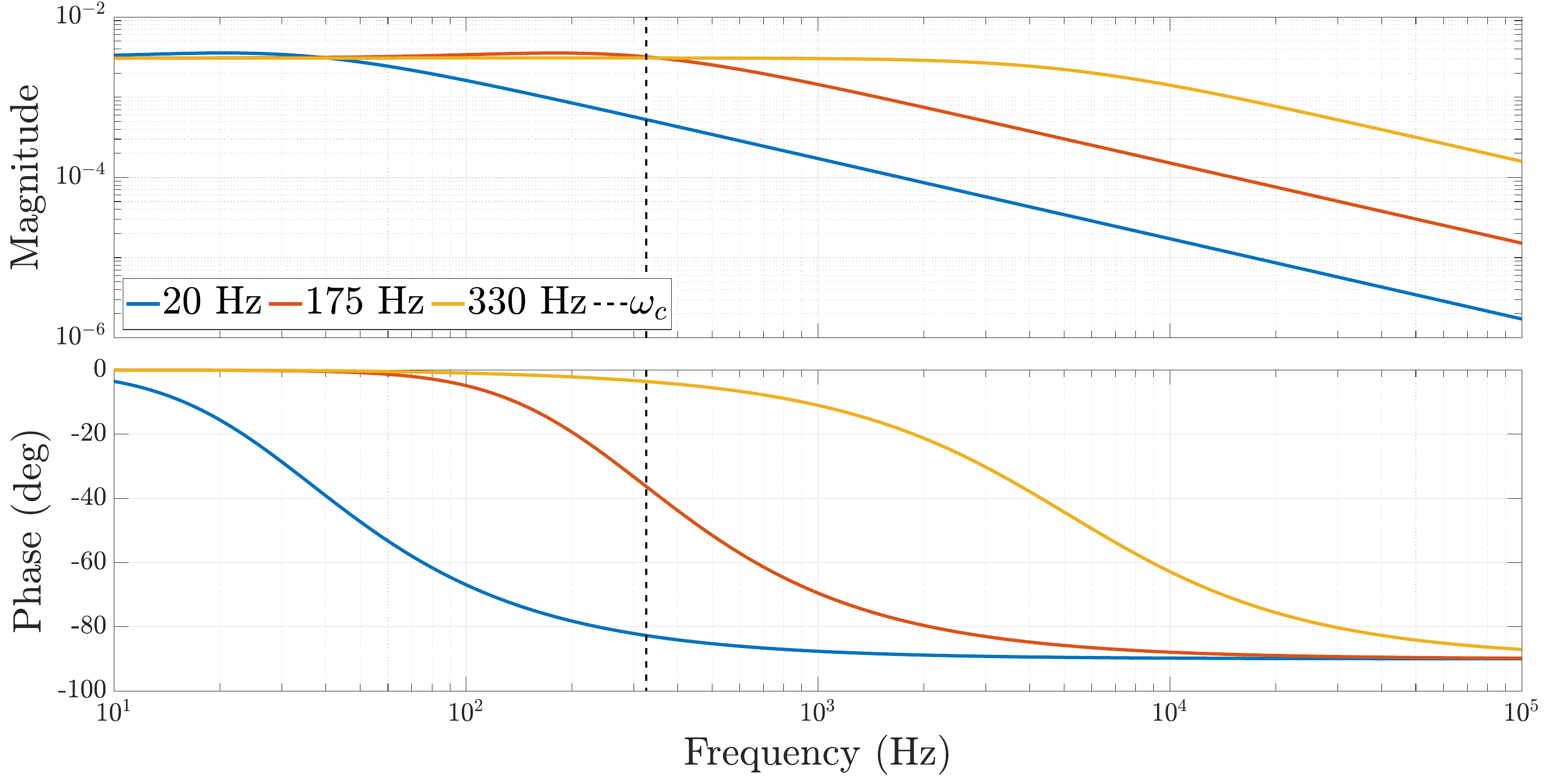}
\caption{Bode diagram representing the frequency response of the PLL.}
\label{fig:PLL_comp}
\end{figure}

In this section, the proposed procedure for assessing the root-cause of instabilities is presented.
First, the grid impedance ($\mathbf{Z}_g(s)$) and the converter admittance ($\mathbf{Y}_c(s)$) are obtained, as shown in Section~\ref{sec:impedance_modelling} and~\ref{sec:admittance_modelling}, respectively.
Although detailed models of the power electronic converter and the grid have been developed to enhance reproducibility, the subsequent analysis relies solely on their frequency responses.

Fig.~\ref{fig:dignity_chart} shows the logarithmic Nyquist plot of $\chi(j\omega)$~\cite{6862243}.
It can be seen that $\chi(j\omega)$ encircles the origin, which means that the interconnected system is unstable. 
The system critical frequency ($\omega_c$) is 490~Hz.
Note that $\chi(j\omega_c)$ is represented as an orange arrow pointing to the left of the complex plane, with a magnitude of 20.2 dB and phase of $\pm 180^\circ$.
In the following lines, a procedure is presented to determine which component contributes more to destabilize the system.
First, the contributions of the admittance terms of the power electronic converter are calculated using (\ref{eqn:ell0})--(\ref{eqn:ell3}), at $j\omega_c$.
These contributions ($\ell_0$, $\ell_1$, $\ell_2$, and $\ell_3$) are displayed in Fig.~\ref{fig:dignity_chart}.
Note that $\ell_0$ and $\ell_3$ have the largest magnitude values.
Meanwhile, the magnitude of $\ell_1$ is very small and cannot be properly seen in the diagram. 
The terms $\ell_0$ and $\ell_3$ are pointing to the left-hand side of the complex plane, and therefore deteriorate the system stability properties. 
In other words, $Y_0(j\omega_c)$ and $Y_3(j\omega_c)$ are deteriorating stability, while $Y_1(j\omega_c)$ and $Y_2(j\omega_c)$ is improving it.
The same procedure can be applied to $\mathbf{Z}_g(s)$, and it is determined that $Z_2(j\omega_c)$ is deteriorating stability.

At this point, it is known that $Y_0(j\omega_c)$, $Y_3(j\omega_c)$ and $Z_2(j\omega_c)$ are deteriorating the system stability.
Indeed, the pairs that are negatively interacting are $Z_2Y_0$ and $Z_2Y_3$.
The significant presence of odd components ($Y_0$ and $Y_3$) on the converter side suggests that the dynamics of asymmetrical control loops are involved.
Therefore, the instability might be related with synchronization loops or DC voltage controllers, among others.
On the grid side, the presence of even components ($Z_2$) is related with passive elements. 
This is expected since the grid consists of inductors and capacitors.
Finally, it is important to mention that all the conclusions drawn in this section are based only on frequency responses.
No analytical models are required until this point.

In this part, it will be confirmed which control block of the power converter is negatively affecting stability.
Therefore, the analytical model of the converter will be used (see Section~\ref{sec:admittance_modelling}), whereas only the frequency response of the grid model is needed. 
The following analysis is not intended to be a step-by-step guide on how to find root causes of instabilities in any scenario.
Rather, it is a specific example on how to use the Pauli decomposition to help finding these root causes.
More systematic approaches are of interest for further research.
For the sake of brevity, only the effect of $Y_0(j\omega_c)$ will be analysed.

First, observing the expression of $\mathbf{Y}_c$ in (\ref{eqn:yc_sum}), it can be seen that $\mathbf{Y}_c$ consists of four terms, added.
Therefore, $Y_0(j\omega_c)$ represents the contribution of the zero (0) components of all admittances.
Fig.~\ref{fig:y0_components} shows a linear phasor diagram of the zero components that form $Y_0$ (green arrow).
The objective is to modify $Y_0(j\omega_c)$ so that the system becomes stable.
Note that, if the magnitude of the zero components of $\mathbf{Y}_{ff}$, $\mathbf{Y}_o$, and $\mathbf{Y}_{cc}$ are reduced, the magnitude of $Y_0$ will also be reduced.
First, the magnitude of $Y_{pll}$ is too low to be considered.
Second, $\mathbf{Y}_{ff}$ and $\mathbf{Y}_{o}$, depend on the converter parameters such as the time delay (\ref{eqn:gff}) and the filter values (\ref{eqn:yc_sum}).
Then, only $\mathbf{Y}_{cc}$ can be modified, since it mainly depends on the design of the current controller and the PLL (\ref{eqn:gcc}).

Fig.~\ref{fig:cc0_components} shows the magnitude and phase of the components of $Y_{cc-0}$.
In this type of representation, magnitude and phases can be simply added.
Among the terms that conform $Y_{cc-0}$, only $H_{cc}(s)$ and $H_{pll}(s)$ have certain degrees of freedom.
Meanwhile, the others ($|I_1|$, $e^{-j\omega_c T_d}$, and $Y_{o-0}$) are fixed.
Therefore, a meaningful strategy is to reduce the magnitude of $H_{cc}(j\omega_c)$ or $H_{pll}(j\omega_c)$, as this will also reduce $|Y_{cc-0}(j\omega_c)|$.
One way of reducing $|H_{cc}(j\omega_c)|$ is to decrease $K_{p-cc}$, but this would also worsen the performance of the current controller.
Moreover, the impact of reducing $K_{p-cc}$ is relatively low, $|H_{cc}(j\omega_c)|$ is already quite low at $\omega_c$.
As an alternative option, Fig.~\ref{fig:PLL_comp} shows the effect of reducing the PLL bandwidth.
In this case, $|H_{pll}(j\omega_c)|$ is reduced and, as a consequence, $|Y_{cc-0}(j\omega)|$, $|Y_0(j\omega)|$ and $|\ell_0|$ are also reduced.
This means that this control modification is the most effective one for improving the stability margins.

\begin{figure}[!t]
\centering
\includegraphics[width=0.99\linewidth]{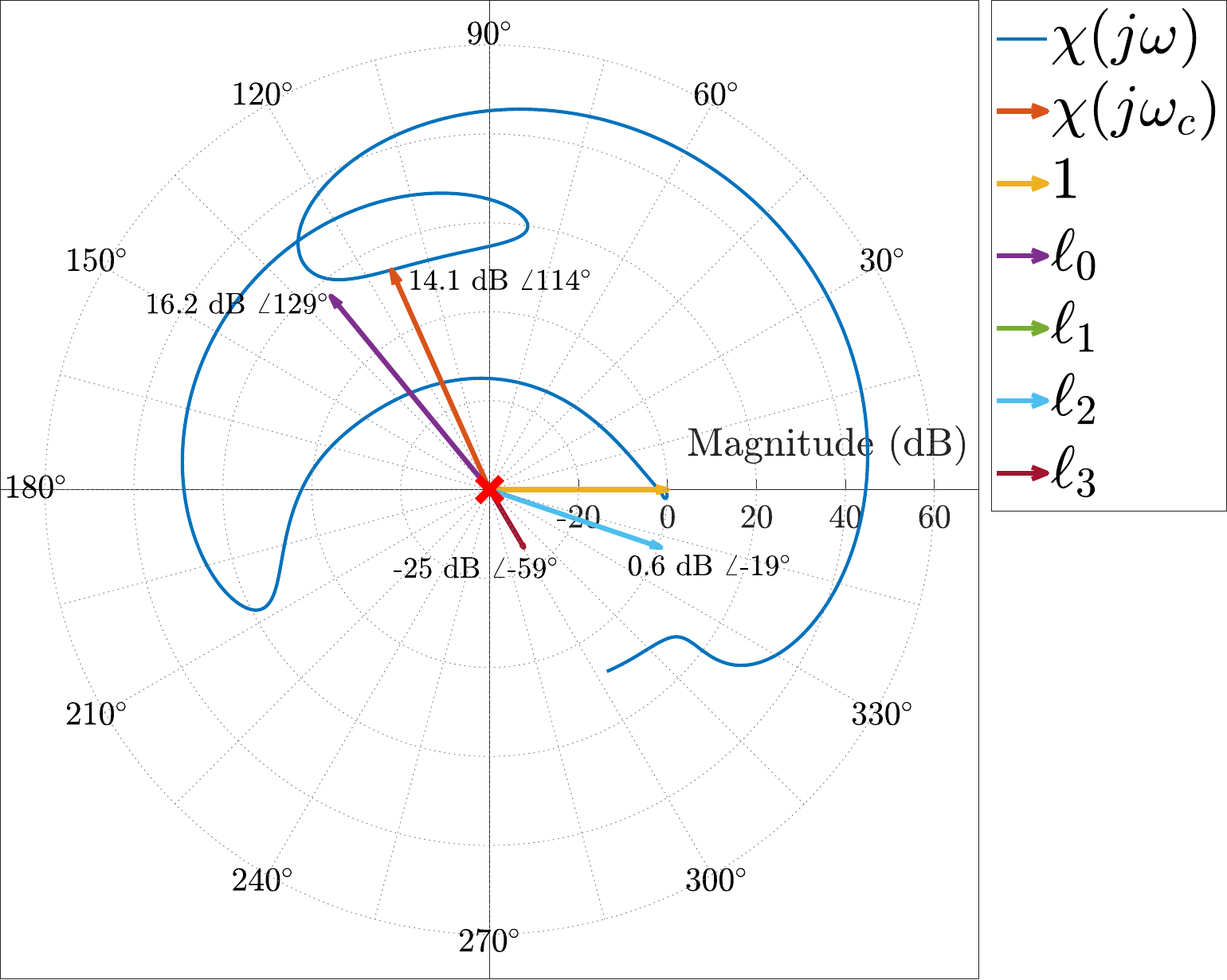}
\caption{Logarithmic Nyquist plot of $\chi(j\omega)$, for the stable case.}
\label{fig:dignity_chart_stable}
\end{figure}

\begin{figure}[!t]
\centering
\includegraphics[width=\linewidth]{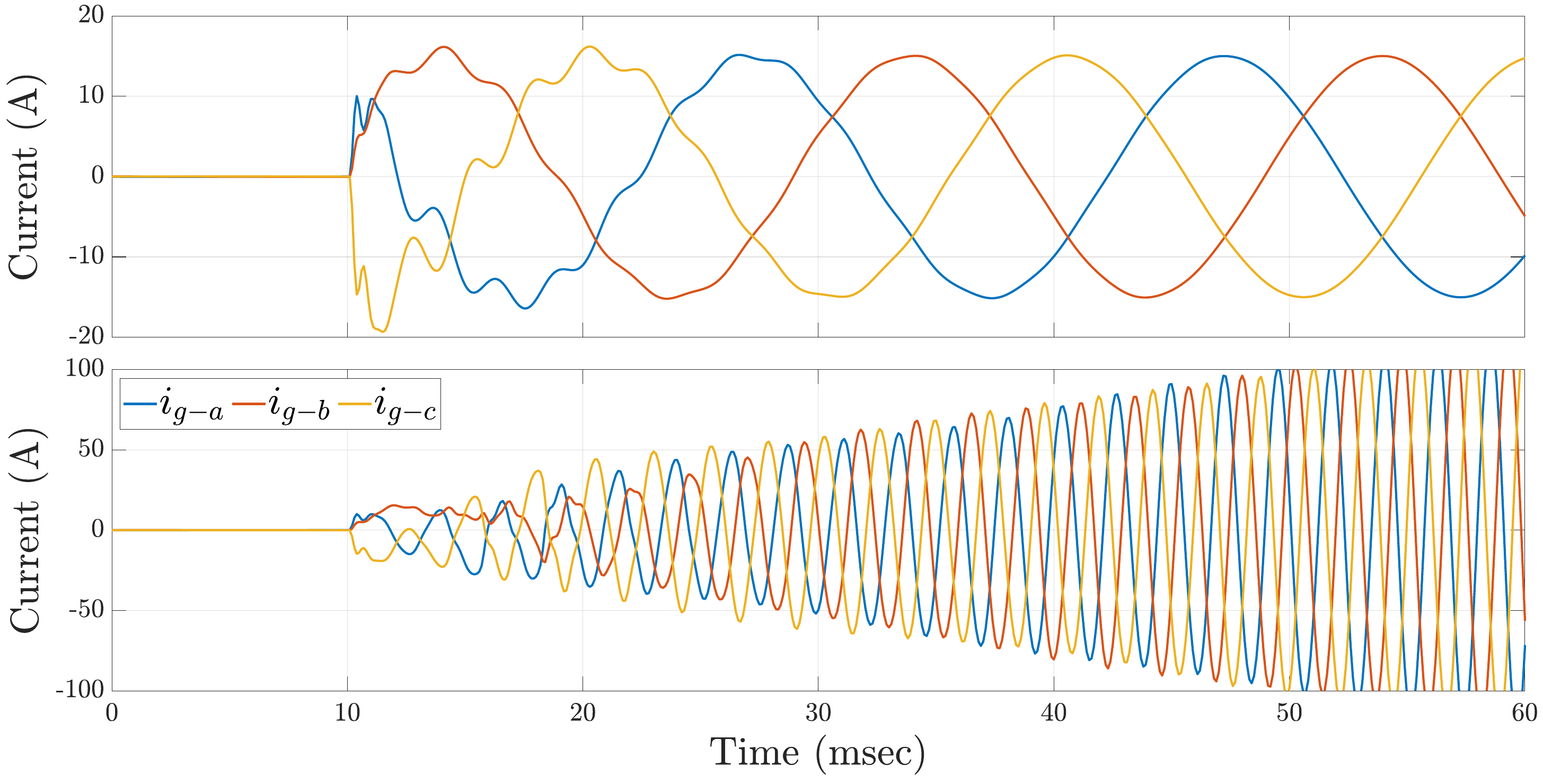}
\caption{Comparison of the resulting currents of the system for the stable (top) and unstable scenario (bottom).}
\label{fig:ig_comp}
\end{figure}

Fig.~\ref{fig:dignity_chart_stable} shows the logarithmic Nyquist plot of the system when the bandwidth of the PLL is reduced to 20~Hz.
In this case, $\chi(j\omega)$ no longer encircles the origin and $\chi(j\omega_c)$ points to the upper-right side of the plot.
It can be seen that now $\ell_0$ points to the upper left side of the plot whereas $\ell_3$ is significantly reduced.
To validate these findings, Fig.~\ref{fig:ig_comp} shows the simulation results of the converter connected to the grid.
In this simulation, the PLL bandwidth is 20~Hz (top) and 330~Hz (bottom).
The rest of parameters have not been modified. 
In the figure, it can be observed that for a PLL bandwidth of 330~Hz, the simulation is unstable. 
However, when the bandwidth is reduced to 20~Hz, the closed-loop system is stable.

\section{Conclusion}
\label{sec:conclusions}
A method to decompose and analyse impedance matrices using Pauli decomposition has been presented in this work.
The proposed method uses Pauli matrices to transform an impedance matrix into a quaternion-like form.
Such representation facilitates the analysis of the root-cause of interactions.

The main advantage of quaternion-like representation is an improved understanding and interpretation of impedance matrices that, as a result, simplifies their mathematical manipulation for stability analyses.
Moreover, it was demonstrated how the influence of decomposed impedance terms are linked to system determinant, traces, characteristic equation, and eigenvalues.

Finally, the principal findings were showcased by using a well-documented benchmark system for impedance based stability analyses.
Using this test case, it was shown firstly that the modelling equations are compact and comprehensible.
The case study was then used to demonstrate how the decomposition separates the interaction terms of the minor loop, thereby enabling a structured interpretation of the components and their contribution to instability.

Future work will focus on application of Pauli matrices to power oscillation damping using grid-connected converter-interfaced generators.
In addition, specific procedures to guarantee the closed-loop stability based on this theory will be developed.

\bibliography{biblio}

@article{Sangwine2011,
author = {Stephen J. Sangwine and Todd A. Ell and Nicolas Le Bihan},
doi = {10.1007/s00006-010-0263-3},
issn = {01887009},
issue = {3},
journal = {Advances in Applied Clifford Algebras},
month = {9},
pages = {607--636},
title = {{Fundamental Representations and Algebraic Properties of Biquaternions or Complexified Quaternions}},
volume = {21},
year = {2011}
}

@article{Wang2018,
author = {Xiongfei Wang and Lennart Harnefors and Frede Blaabjerg},
doi = {10.1109/TPEL.2017.2684906},
issn = {08858993},
issue = {2},
journal = {IEEE Transactions Power Electronics},
month = {2},
pages = {1775--1787},
title = {{Unified Impedance Model of Grid-Connected Voltage-Source Converters}},
volume = {33},
year = {2018}
}

@article{Harnefors2020,
author = {Lennart Harnefors and Xiongfei Wang and Shih Feng Chou and Massimo Bongiorno and Marko Hinkkanen and Mikko Routimo},
issue = {2},
journal = {IEEE Journal of Emerging and Selected Topics in Power Electronics},
month = {6},
pages = {1911--1921},
publisher = {Institute of Electrical and Electronics Engineers Inc.},
title = {{Asymmetric Complex-Vector Models with Application to VSC-Grid Interaction}},
volume = {8},
year = {2020}
}

@techReport{Zhang2017,
author = {Hongbing Zhang},
title = {{Intuitive Geometric Significance of Pauli Matrices and Others in a Plane}},
year = {2017}
}

@inbook{Amin2019,
author = {Mohammad Amin and Chen Zhang and Atle Rygg and Marta Molinas and Eneko Unamuno and Mohamed Belkhayat},
booktitle = {Wiley Encyclopedia of Electrical and Electronics Engineering},
month = {5},
pages = {1--22},
publisher = {Wiley},
title = {{Nyquist Stability Criterion and its Application to Power Electronics Systems}},
year = {2019}
}

@ARTICLE{García2024,
author={Velasco, Manel and Zaplana, Isiah and Dòria-Cerezo, Arnau and Duarte, Josué and Martí, Pau},
journal={IEEE Transactions on Industrial Electronics}, 
title={{Introducing Modeling, Analysis, and Control of Three-Phase Electrical Systems Using Geometric Algebra}}, 
year={2025},
pages={1-10}
}

@article{Rygg2016,
author = {Atle Rygg and Marta Molinas and Chen Zhang and Xu Cai},
issn = {21686785},
issue = {4},
journal = {IEEE Journal Emerging Selected Topics Power Electronics},
month = {12},
title = {{A Modified Sequence-Domain Impedance Definition and Its Equivalence to the dq-Domain Impedance Definition for the Stability Analysis of AC Power Electronic Systems}},
year = {2016}
}

@article{Zhu2023,
author = {Yue Zhu and Yunjie Gu and Yitong Li and Timothy C. Green},
issue = {2},
journal = {IEEE Transactions on Power Systems},
month = {3},
pages = {1642--1654},
title = {{Impedance-Based Root-Cause Analysis: Comparative Study of Impedance Models and Calculation of Eigenvalue Sensitivity}},
volume = {38},
year = {2023}
}

@article{Zhu2022,
author = {Yue Zhu and Yunjie Gu and Yitong Li and Timothy Green},
issn = {15580679},
issue = {1},
journal = {IEEE Transactions on Power Systems},
month = {1},
pages = {343--353},
title = {{Participation Analysis in Impedance Models: The Grey-Box Approach for Power System Stability}},
volume = {37},
year = {2022}
}

@article{Li2021,
author = {Yitong Li and Yunjie Gu and Yue Zhu and Adria Junyent-Ferre and Xin Xiang and Timothy C. Green},
issn = {19410107},
issue = {3},
journal = {IEEE Transactions on Power Electronics},
month = {3},
pages = {3377--3395},
title = {{Impedance Circuit Model of Grid-Forming Inverter: Visualizing Control Algorithms as Circuit Elements}},
volume = {36},
year = {2021}
}

@article{DionysiosMoutevelis2024,
author = {Dionysios Moutevelis and Javier Roldan-Perez and Pablo Rodríguez-Ortega and Milan Prodanovic},
issn = {02137585},
issue = {2},
journal = {IEEE Transactions on Energy Conversion},
month = {9},
pages = {187--214},
title = {{Virtual Synchronous Machine Design for Islanded Microgrids Using the Extended Impedance Criterion with Grid Frequency Dynamics Included}},
volume = {125},
year = {2024}
}

@article{Li2020,
author = {Chuanyue Li and Jun Liang and Liana M. Cipcigan and Wenlong Ming and Frederic Colas and Xavier Guillaud},
issn = {15580059},
issue = {4},
journal = {IEEE Transactions Energy Conversion},
month = {12},
title = {{DQ Impedance Stability Analysis for the Power-Controlled Grid-Connected Inverter}},
year = {2020}
}

@article{Hitzer2024,
author = {Eckhard Hitzer and Manos Kamarianakis and George Papagiannakis and Petr Vašík},
issn = {10991476},
issue = {14},
journal = {Mathematical Methods in the Applied Sciences},
month = {9},
pages = {11368--11384},
publisher = {John Wiley and Sons Ltd},
title = {{Survey of New Applications of Geometric Algebra}},
volume = {47},
year = {2024}
}

@article{Sun2011,
author = {Jian Sun},
issn = {08858993},
issue = {11},
journal = {IEEE Transactions on Power Electronics},
pages = {3075--3078},
title = {{Impedance-Based Stability Criterion for Grid-Connected Inverters}},
volume = {26},
year = {2011}
}

@article{Zhang2020,
author = {Chen Zhang and Marta Molinas and Atle Rygg and Xu Cai},
doi = {10.1109/JESTPE.2019.2914560},
issn = {21686785},
issue = {3},
journal = {IEEE Journal of Emerging and Selected Topics in Power Electronics},
month = {9},
pages = {2520--2533},
title = {{Impedance-Based Analysis of Interconnected Power Electronics Systems: Impedance Network Modeling and Comparative Studies of Stability Criteria}},
volume = {8},
year = {2020}
}

@inproceedings{Fang2024,
author = {Jingyang Fang},
doi = {10.1109/ECCE55643.2024.10861569},
isbn = {9798350376067},
booktitle = {2024 IEEE Energy Conversion Congress and Exposition, ECCE 2024 - Proceedings},
pages = {1237-1244},
title = {{Three-Phase Circuit Analysis through Quaternions and Biquaternions}},
year = {2024}
}

@article{Tudor2010,
author = {Tiberiu Tudor},
issn = {00304026},
issue = {13},
journal = {Optik},
pages = {1226--1235},
publisher = {Urban und Fischer Verlag Jena},
title = {{Vectorial Pauli Algebraic Approach in Polarization Optics. I. Device and State Operators}},
volume = {121},
year = {2010}
}

@book{kundur1994power,
  title={Power System Stability and Control},
  author={Kundur, P. and Balu, N.J. and Lauby, M.G.},
  isbn={9780070359581},
  lccn={93021456},
  series={EPRI power system engineering series},
  year={1994},
  publisher={McGraw-Hill Education}
}

@ARTICLE{MCPE_SmallSignal,
author={Chen, Qifan and Bu, Siqi and Chung, Chi Yung},
journal={Journal of Modern Power Systems and Clean Energy}, 
title={{Small-Signal Stability Criteria in Power Electronics-Dominated Power Systems: A Comparative Review}}, 
year={2024},
volume={12},
number={4},
pages={1003--1018}
}

@article{Pedra2024,
   author = {J. Pedra and L. Sainz and Ll Monjo},
   issn = {19374208},
   issue = {2},
   journal = {IEEE Transactions on Power Delivery},
   month = {4},
   pages = {1283-1298},
   title = {{Review and Improvements to the Measurements of the VSC Impedance Transfer Matrix}},
   volume = {39},
   year = {2024}
}

@article{Fan2020,
   author = {Lingling Fan and Zhixin Miao},
   issn = {15580679},
   issue = {4},
   journal = {IEEE Transactions on Power Systems},
   month = {7},
   pages = {3312-3315},
   title = {{Admittance-Based Stability Analysis: Bode Plots, Nyquist Diagrams or Eigenvalue Analysis?}},
   volume = {35},
   year = {2020}
}

@article{SU2020475,
title = {{Fast Frequency Response of Inverter-Based Resources and Its Impact on System Frequency characteristics}},
journal = {Global Energy Interconnection},
volume = {3},
number = {5},
pages = {475-485},
year = {2020},
issn = {2096-5117},
author = {Lining Su and Xiaohui Qin and Shang Zhang and Yantao Zhang and Yilang Jiang and Yi Han},
}

@article{Hatziargyriou2021,
   author = {Nikos Hatziargyriou and Jovica Milanovic and Claudia Rahmann and Venkataramana Ajjarapu and Claudio Canizares and Istvan Erlich and David Hill and Ian Hiskens and Innocent Kamwa and Bikash Pal and Pouyan Pourbeik and Juan Sanchez-Gasca and Aleksandar Stankovic and Thierry Van Cutsem and Vijay Vittal and Costas Vournas},
   issn = {15580679},
   issue = {4},
   journal = {IEEE Transactions on Power Systems},
   month = {7},
   pages = {3271-3281},
   title = {{Definition and Classification of Power System Stability - Revisited \& Extended}},
   volume = {36},
   year = {2021}
}

@inproceedings{cheng2021ibr,
  title     = {{IBR Grid Integration: Operation Challenges and the New Stability Assessment Paradigm Based on Black-box Models}},
  author    = {Cheng, Y. and Fan, L.},
  booktitle = {IEEE Power \& Energy Society General Meeting (PESGM)},
  year      = {2021},
}

@INPROCEEDINGS{6862243,
author={Zanasi, Roberto and Grossi, Federica},
booktitle={2014 European Control Conference (ECC)}, 
title={Open and Closed Logarithmic Nyquist Plots}, 
year={2014},
volume={},
number={},
pages={850-855}
}

@ARTICLE{MentiTCAS,
author={Menti, Anthoula and Zacharias, Thomas and Milias-Argitis, John},
journal={IEEE Transactions on Circuits and Systems I: Regular Papers}, 
title={{Geometric Algebra: A Powerful Tool for Representing Power Under Nonsinusoidal Conditions}}, 
year={2007},
volume={54},
number={3},
pages={601-609},
doi={10.1109/TCSI.2006.887608}
}

@article{FGIL_GeometricAlgebra,
author = {Montoya, Francisco G. and Eid, Ahmad H.},
title = {{Formulating the Geometric Foundation of Clarke, Park, and FBD Transformations by Means of Clifford's Geometric Algebra}},
journal = {Mathematical Methods in the Applied Sciences},
volume = {45},
number = {8},
pages = {4252--4277},
year = {2022}
}

@ARTICLE{CastillaGA,
author={Castilla, M. and Bravo, Juan Carlos and Ordonez, M. and Montano, Juan Carlos},
journal={IEEE Transactions on Circuits and Systems I: Regular Papers}, 
title={{Clifford Theory: A Geometrical Interpretation of Multivectorial Apparent Power}}, 
year={2008},
volume={55},
number={10},
pages={3358-3367},
}

@ARTICLE{MiltonGA,
author={Castro-Nunez, Milton and Castro-Puche, Róbinson},
journal={IEEE Transactions on Circuits and Systems I: Regular Papers}, 
title={{Advantages of Geometric Algebra Over Complex Numbers in the Analysis of Networks With Nonsinusoidal Sources and Linear Loads}}, 
year={2012},
volume={59},
number={9},
pages={2056-2064},
}

@ARTICLE{10262032,
  author={Wu, Heng and Zhao, Fangzhou and Wang, Xiongfei},
  journal={IEEE Power Electronics Magazine}, 
  title={A Survey on Impedance-Based Dynamics Analysis Method for Inverter-Based Resources}, 
  year={2023},
  volume={10},
  number={3},
  pages={43-51},
  doi={10.1109/MPEL.2023.3303104}
}

@software{pauliCode,
  author  = {Josué Andino and Milan Prodanovic and Javier Roldán-Pérez},
  title   = {{MATLAB Tool for performing Pauli Decomposition of Impedance Matrices}},
  doi     = {10.5281/zenodo.17829032},
  url     = {https://doi.org/10.5281/zenodo.17829032}
}

\appendices 

\section{Mathematical Tools}
This section briefly summarises the mathematical tools that are required to follow this work.
Section~\ref{sec:reim} describes the real and imaginary operations using Pauli matrices.
Finally, Section~\ref{sec:rel} describes the relationship between the impedance quaternion in $dq$ coordinates, $pn$ sequence, and $\alpha\beta$ frame.

\subsection{Projection Operators}
\label{sec:reim}
The $d$ and $q$ components of a complex variable can be obtained by using the following expressions:
\begin{align}
\Re{i_{dq}} 
= 
\frac{ \mathbf{i}_{dq} 
+ 
\mathbf{i}_{dq}^* }{2}
\quad \text{and} \quad
\Im{i_{dq}} 
= 
\frac{ \mathbf{i}_{dq} 
- 
\mathbf{i}_{dq}^* }{2j},
\end{align}
where $\Re{\cdot}$ and $\Im{\cdot}$ are functions that extract the real and imaginary components of a complex variable, respectively.
These two operators can also be written in matrix form using the Pauli decomposition method.
For that, it is required to define the projectors $P_d$ and $P_q$, as:
\begin{equation}
P_d := \frac{I+K}{2} \quad \text{and} \quad P_q := \frac{I-K}{2}.
\end{equation}
To extract the $d$ and $q$ components, the following operation is performed:
\begin{equation}
\mqty[ i_d \\ 0 ] 
= 
P_d \mathbf{i}_{dq}
\quad
\text{and}
\quad
\mqty[ 0 \\ i_q ] 
= 
P_q \mathbf{i}_{dq}.
\end{equation}
It should be noted that these projectors are linear operators in the Pauli representation.
However, the original real and imaginary operators for complex variables are non-linear.

\subsection{Other Representations of the Impedance Matrix }
\label{sec:rel}

In the literature, several other representations of the impedance matrix can be found.
Among them, the positive-negative sequence ($\mathbf{Z}_{pn}(s)$) is commonly used.
In this case, a $2x2$ matrix made of four complex transfer functions is used, and it has the following relationship with the $dq$ impedance matrix:
\begin{align}
\mathbf{Z}_{pn}(s)
=
A \mathbf{Z}_{dq}(s) A^{-1}
=
\mqty[ Z_0+jZ_2 & Z_3+jZ_1 \\ 
Z_3-jZ_1 & Z_0-jZ_2 ],
\end{align}
with
\begin{align}
A 
= 
\mqty[ 1 & \phantom{-}j \\ 
1 & -j ].
\end{align}
The matrices $J' = A J A^{-1}$ and $K' = A K A^{-1}$ can be defined, and then $\mathbf{Z}_{pn}(s)$ can be written as:
\begin{align}
\label{eqn:Zpn}
\mathbf{Z}_{pn}(s)
&=
Z_0(s)I + Z_2(s)J' + Z_3(s)K' + Z_1(s)J'K'.
\end{align}
It can be seen that (\ref{eqn:Zpn}) can be decomposed using Pauli matrices, as the $dq$ impedance matrix in (\ref{eqn:pauli_impedance}).
Moreover, the matrices $J'$ and $K'$ have the same properties compared to the original matrices ($J$ and $K$), which described in Section~\ref{sec:quaternion}.
Finally, the impedance in $\alpha\beta$ ($\mathbf{Z}_{\alpha\beta}(s)$) also has the same structure, since it is a version of $\mathbf{Z}_{pn}(s)$, but including a frequency shift:
\begin{align}
\mathbf{Z}_{\alpha\beta}(s) 
= 
\mathbf{Z}_{pn}(s-j\omega_1).
\end{align}

\end{document}